\documentclass[sigconf,screen]{acmart}
\usepackage{array}
\usepackage{colortbl}

\AtBeginDocument{%
  \providecommand\BibTeX{{%
    \normalfont B\kern-0.5em{\scshape i\kern-0.25em b}\kern-0.8em\TeX}}}

\copyrightyear{2022}
\acmYear{2022}
\setcopyright{acmlicensed}\acmConference[COMPUTE 2022]{COMPUTE 2022}{November 9--11, 2022}{Jaipur, India}
\acmBooktitle{COMPUTE 2022 (COMPUTE 2022), November 9--11, 2022, Jaipur, India}
\acmPrice{15.00}
\acmDOI{10.1145/3561833.3561842}
\acmISBN{978-1-4503-9775-9/22/11}

\acmConference[COMPUTE '22]{}{November 09--11, 2022}{Jaipur, India}
\begin{document}

\title{A study of the design and documentation skills of industry-ready CS students*}

\author{Mrityunjay Kumar}
\affiliation{%
  \institution{IIIT Hyderabad}
  \city{Hyderabad}
  \state{Telangana}
  \country{India}
}
\email{mrityunjay.k@research.iiit.ac.in}

\author{Venkatesh Choppella}
\affiliation{%
  \institution{IIIT Hyderabad}
  \city{Hyderabad}
  \state{Telangana}
  \country{India}
}
\email{venkatesh.choppella@iiit.ac.in}

\thanks{*CS undergraduate students who are in fourth year or beyond are referred to as industry-ready in this paper}

\renewcommand{\shortauthors}{Kumar and Choppella}

\begin{abstract}
  An engineer in a product company is expected to design a good solution to a computing problem (Design skill) and articulate the solution well (Expression skill).  We expect an industry-ready student (final year student or a fresh campus hire) as well to demonstrate both these skills when working on simple problems assigned to them. 
  
  This paper reports on the results when we tested a cohort of participants (N=16) for these two skills.  We created two participant groups from two different tiers of college, one from a Tier 1 college (who were taking an advanced elective course), and another from Tier 2 colleges (who had been hired for internship in a SaaS product company).  We gave them a simple design problem and evaluated the quality of their design and expression.  Design quality was evaluated along three design principles of Abstraction, Decomposition, and Precision (adapted from the Software Engineering Book of Knowledge).  Expression quality was evaluated using criteria we developed for our study that is based on the diversity and density of the expressions used in the articulation.  
  
  We found the students lacking in design and expression skills. Specifically, a) they struggled with abstraction as a design principle, b) they did not use enough modes of expressions to articulate their design, and c) they did not use enough formal notations (UML, equations, relations, etc.). We also found significant difference in the performance between the two participant groups. 
\end{abstract}



\maketitle

\section{Introduction}
\label{sec:intro}
The Indian product startup ecosystem continues to grow.  In 2021 alone, there were more than 2200 startups founded, and tech startups raised more than 24B USD \cite{nasscom_zinnov_2021}, including many SaaS startups \cite{saasboomi_2021}.  These companies require a strong pool of software engineers and industry-ready students.  However, the quality of students graduating from engineering colleges is poor; less than 4\% of the engineers have enough skills for these startups and product companies \cite{employability_2019,blom2011employability}.

Design is a key step in an engineering process \cite{taylor2007software}, and the ability to design is a key skill for a software engineer.  This translates into three competencies that companies seek: 1) ability to design solutions to problems, 2) ability to articulate the design, and 3) collaborate with the team on the design and make it fit for the company. \cite{10.1145/1404520.1404522,10.1145/1734263.1734359}. 

Given the need for skilled engineers by startups and product companies, and given that design ability is a key skill, we want to understand the level of design skills in industry-ready students (CS undergraduate students who are in fourth year or beyond) of the engineering colleges. 

SWEBOK \cite{borque2014guide} lists seven software design principles: 1) Abstraction, 2) Coupling and Cohesion, 3) Decomposition and Modularization, 4) Encapsulation and Information Hiding, 5) Separation of interface and implementation, 6) Sufficiency, Completeness, and Primitiveness, and 7) Separation of concerns. To restrict our scope, we selected three principles from this list to evaluate the students on.  
\textbf{Decomposition} reflects the ability to break down a large problem into several small ones that can be independently solved and composed into a solution for the large problem \cite{10.1145/1960275.1960307}. \textbf{Abstraction} \cite{liskov1986abstraction} is the ability to generalize and focus on essential information, either through parameterization or specification. \textbf{Precision} (Sufficiency or Completeness) is the ability to handle all stated requirements correctly with the proposed design. 

For our experiment, we define Design Skill as follows: 

  \textbf{Design Skill}: \textit{Ability to apply key design principles of Abstraction, Decomposition and Precision in a solution design}'. 
  
This paper presents our findings from an experiment we ran with two participant groups to understand their Design Skill. 

We evaluated the participant submissions on two dimensions: a) Design Quality, which evaluated the design against the three design principles, and b) Expression Quality, which evaluated the design against the usage of different ways of expressing their design - text, pictures, notations, etc. 
  
\textit{Research Questions}. We proceeded to answer three research questions about the design produced by the participants: 
  \begin{quote}
  \textit{RQ1: Do industry-ready students produce good quality design?}
  
  \textit{RQ2: Do industry-ready students articulate their design well?}

  \textit{RQ3: Is there a difference in performance between Tier 1 and Tier 2 college students?}
  \end{quote}


  The rest of the paper is organized as follows: Section \ref{sec:related} presents related work on design skills. Section \ref{sec:exp-setup} describes the design of the experiment. Section \ref{sec:eval-setup} lays out the evaluation criteria and other details about the evaluation. Section \ref{sec:findings} discusses the findings based on the analysis of the results. Finally, Section \ref{sec:conclusions} presents our conclusion and plan for future work. 
\section{Related Work}
\label{sec:related}
The software industry is a knowledge-intensive industry. As Robillard \cite{10.1145/291469.291476} says, '\textit{Software development is the processing of knowledge in a very focused way}'. This processing applies the knowledge to solve the computational problem at hand, and such skills are required for industry-ready students to do well in their job. A precise definition of what is software design may be elusive \cite{viviani2022really}, but their importance is well-established \cite{taylor2007software}. Hewner and Guzdial  \cite{10.1145/1734263.1734359} showed, albeit in a limited game company setting, that "\textit{Being able to solve algorithmically challenging problems}" and "\textit{Ability to build a good object design for a large system and understand the implications}" are held as very important skills for the industry. Radermacher et al \cite{10.1145/2591062.2591159} showed that '\textit{Problem solving}' is in top 5 knowledge competencies identified in recently graduated students. Hartman et al \cite{10.1145/319059.323455} mention '\textit{The two most important skills which a student embarking on his career can have are communication skills and problem solving skills.}'

Of the seven design principles laid out in SWEBOK \cite{borque2014guide} we chose Abstraction, Decomposition, and Sufficiency as key skills to evaluate for these industry-ready students. 

\textbf{Abstraction}. Kramer \cite{10.1145/1232743.1232745} strongly argues that Abstraction is key to computing. Liskov \cite{liskov1986abstraction,liskov2000program} and others \cite{10.1145/1370164.1370174} have suggested many approaches to achieve abstraction in the design (abstraction by parameterization and by specification are two key ones) and we expect the students to demonstrate this ability. 

\textbf{Decomposition}. Hoek and Lopez \cite{10.1145/1960275.1960307} claim that "..Without [Modularization], large software systems simply could not be realized". Various decomposition and modeling techniques (functional, subsystem-based, abstraction, etc.) have been in use and can be applied when one designs a system \cite{10.1145/2016039.2016104,10.1007/s00165-017-0428-0}, and we expect the industry-ready students to evidence modularity and decomposition in their design. 

\textbf{Precision}. Sufficiency and completeness of the design ensures there is less rework required after release. Precision can be usually improved by leveraging formal visual languages like UML \cite{booch2005unified}, or algebraic-logical languages like Z notation \cite{spivey1988understanding}. We expect the industry-ready students to use some of the tools to achieve precision without sacrificing abstraction. 

\section{Experiment Setup}
\label{sec:exp-setup}
\subsection{Study Design}
\label{subsec:studydesign}
We asked students to devise a design for their solution for a given problem and then describe their design in 2-3 pages. The problem was given as a take-home assignment to two groups of students with 7 days to complete it, so there was enough time given for the students to do a good job. 
We wanted to select a simple problem that allows the elements of abstraction, decomposition, and precision to a level we expect an industry-ready student to possess. We selected a board game (Tic-tac-toe) that is usually used as a non-trivial programming problem for students to tackle as a beginner programmer. While it is hard to show any selected problem is the best choice, Tic-tac-toe solution design demonstrates abstraction, decomposition, and precision well (see the solution outline \href{https://bit.ly/compute22sol}{bit.ly/compute22sol}) so we believe it is a good choice of a design problem for the students. 

\textit{Problem summary: } We asked them to consider the 2-person game tic-tac-toe in which a human plays the game with a computer. They were then asked to think about the design of a software solution for this and then describe the design in writing, with enough precision that a first year CS student can code against. We also asked them not to write code or pseudo-code and stay at the design level.
Complete problem text that was given to them: \href{https://bit.ly/compute22prob}{bit.ly/compute22prob}

\subsection{Participants}
We selected two types of participants for this experiment. 
\textbf{Group A} consisted of third- and fourth-year students of a Tier 1 college who were enrolled for a "Topics in Software Foundations" course (Computer Science Elective). This design problem was given as part of their graded assignments. 
\textbf{Group B} consisted of students from good Tier 2 college who had gone through two rounds of interview for internship at a local SaaS (unicorn) company and were selected for internship.    

\textit{Incentive for good work}: To ensure that students and interns do their best on these problems, this was given as part of the graded assignment to Group A participants. The first author delivered a Design Masterclass for Group B participants as part of their training and this design problem was given as pre-requisite for attending this workshop. 

\section{Evaluation Setup}
\label{sec:eval-setup}
As mentioned in the Introduction, we evaluated the submissions on two dimensions: Design Quality and Expression Quality.  Participants submitted their work as PDF documents that captured their design, and we used expert evaluation (by author) of the submissions to generate the score for analysis.

\subsection{Design Quality Evaluation}
\label{subsec:design-eval}
The submissions were evaluated on the three selected design principles:
\begin{enumerate}
    \item \textbf{Abstraction}
    \item \textbf{Decomposition}
    \item \textbf{Precision} 
\end{enumerate}

We then created a rubric to evaluate the evidence of each of these design principles in the submission, using the solution outline created earlier (\href{https://bit.ly/compute22sol}{bit.ly/compute22sol}).

\subsubsection{Decomposition rubric}
\label{subsubsec:decomposition}
\textit{Rubric:} We evaluate based on how many of the components (as identified in solution outline or similar) are identified as part of decomposition and used the score of 0 (No components identified), 1 (one or two components identified), and 2 (three or more components identified). 

\subsubsection{Abstraction rubric}
\label{subsubsec:abstraction}
According to Liskov \cite{liskov1986abstraction,liskov2000program}, abstraction can be achieved in two ways: a) \textit{Abstraction by Parameterization} - The system should demonstrate that general-purpose functions, procedures, and services have been designed and used, and b) \textit{Abstraction by Specification} - The system should demonstrate that services and functions are defined such that interface definitions are enough to use them.

Abstraction is a 'good' way to decompose - so if decomposition is not done well (say only one component is identified), abstraction will not be in evidence.  To ensure we do not penalize twice, we only look for abstraction in the defined components and give full marks if it is done well (even when the remaining components are missing). 
We also want to ensure that the students do not resort to low level details like code or pseudo-code and so we penalize it. 

\textit{Rubric:} We evaluate based on evidence of abstraction and use scores 0 (No abstraction evidenced), 1 (1-2 methods identified for the components and/or some interface definitions but mostly incomplete), and 2 (Complete interface definitions for components, the methods on the components are sufficiently specified or described). 

\subsubsection{Precision rubric}
\label{subsubsec:precision}
We listed down key scenarios and evaluated the submissions against these: a) The system declares a result correctly in all cases (win or draw), b) The system has a move for every valid move from the human player, c) The system's move is always the most optimal in that situation, d) The board state is always available, and e) The board state is available correctly after every move. 

Result of an evaluation of a scenario can be Pass (the description can correctly handle the scenario), Inconclusive (not enough information to say pass or not) or Fail (incorrect handling of the scenario by the description).

\textit{Rubric:} We evaluate based on how many scenarios Pass, and use scores of 0 (None of the scenarios pass), 1 (50\% of the scenarios pass), and 2 (All scenarios pass).

\subsubsection{Scoring}
Design quality was scored using the rubrics for Abstraction (section \ref{subsubsec:abstraction}, Decomposition (section \ref{subsubsec:decomposition}) and Precision (section \ref{subsubsec:precision}) 
To reflect the relative prioritization across these three parameters, we assigned different weights to the parameters:  Decomposition, Abstraction, and Precision were Fibonacci-weighted (2, 3, 5, respectively) to reflect the fact that abstraction is more important than just decomposition, and precision is more important than both.   Average scores within each parameter were normalized to 100\% to compare them across groups and across parameters.  Table \ref{table:score} presents the scores and averages for the participants and the groups. 

We also calculated the pair-wise correlation among the Decomposition, Abstraction and Precision scores of the Group A participants (Table \ref{table:corrA}).  This was not done for Group B participants, since Precision scores for all samples were 0. 

\subsection{Expression Quality Evaluation}
\label{subsec:exp-eval}
An expression is a linguistic or domain-specific means to describe something (in this case, design) - like text, diagrams, flow charts, UML, etc.  This is also called mode, and social scientists and linguists refer to it as multimodal text (we use 'expression' in this paper). 

\begin{figure}
    \centering
    \includegraphics[width=\columnwidth]{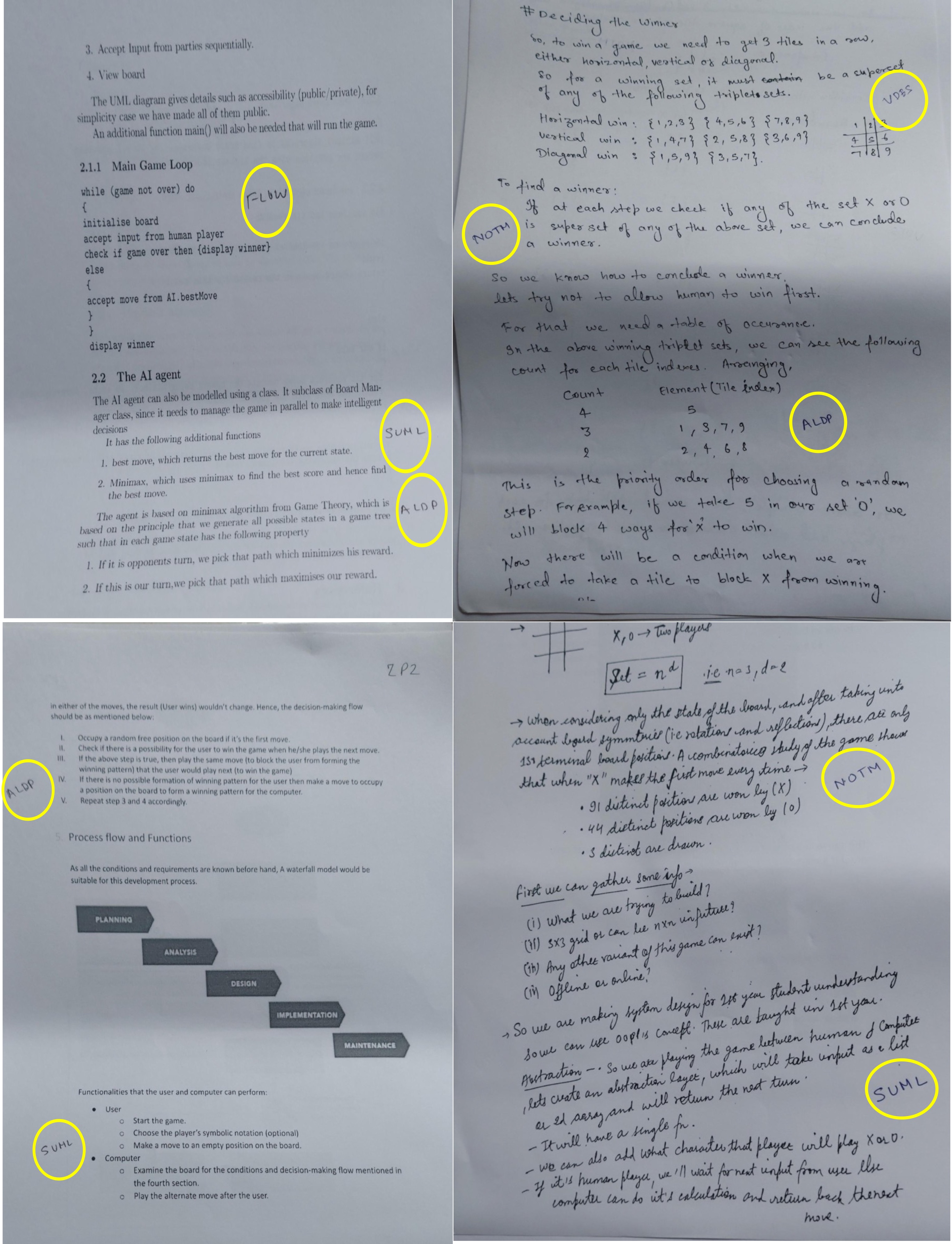}
    \caption{Encoded submission samples (codes are yellow-circled)}
    \label{fig:enc1}
\end{figure}

We encoded the submission to identify the expression types they used. The submission was broken into chunks (a text paragraph or a picture) and each chunk was encoded using one of the codes defined (section \ref{subsubsec:expr-types}).  See Figure \ref{fig:enc1} for some samples of encoding of the submissions. 
We generated two metrics from the encoding. 
\begin{itemize}
    \item \textbf{Diversity} measures how many distinct expression types were used
    \item \textbf{Density} counts total number of expressions used irrespective of type. 
\end{itemize}
These two metrics were reported for each student (Table \ref{table:score}).  Technically, this measures quantity and not quality.  However, it has been shown that using multimodal texts (expressions) improve the quality of the text \cite{ajayi2009english,bezemer2008writing}.  For this paper, we have used these two metrics as proxy to quality of their expression.  We expect that every participant uses each of these expression types at least once, which translates to a Diversity average of 7 (everyone uses all types).    

\subsubsection{Expression types}
\label{subsubsec:expr-types}

This is the list of the expressions we looked for. The 4-letter code is the representation during encoding. 
\begin{itemize}
    \item Visual Description, Diagrams (VDES)
    \item Procedures and Interface definitions (PROC)
    \item Pre-defined abstractions like algorithms and design patterns (ALDP)
    \item Flow Chart or steps description (FLOW)
    \item Static modeling diagrams - UML or similar - for Class, Object, Component depiction etc. (SUML)
    \item Behavioral or Dynamic modeling diagrams - UML or similar - for Activity, Sequence, States etc. (BUML)
    \item Modeling and Specifications through notations like Z-notation, Sets/relations, formalized English, etc. (NOTM)
\end{itemize}

\section{Findings and Discussion}
\label{sec:findings}
(Abbreviations used in this section. \textit{\textbf{DC}: Decomposition, \textbf{AB}: Abstraction, \textbf{PR}: Precision, \textbf{CWS}: Composite Weighted Score, \textbf{DV}: Expression Diversity, \textbf{DN}: Expression Density, \textbf{M}: Mean, \textbf{SD}: Standard Deviation, \textbf{CoV}: Coefficient of Variance})

We set out to understand the Design Quality (RQ1) and Expression Quality (RQ2) of the participants, and whether there is a difference in performance between the two groups (RQ3).

Mean Composite Weighted Score for the two groups (together) is 3.06 (out of 10), which is quite low (barely passing marks!). This suggests that \textbf{Design Quality for this cohort is poor}. We find that students used a small number of expression types (DV Mean: 2.69) and number of expressions (DN average: 4.69)  when articulating their design. We expected all types to be used and many more expressions to be used on an average, so \textbf{Expression Quality is poor as well}. 
We expected the groups to have similar performance. However that is not the case; \textbf{there is a difference in performance between Group A and B}. 
We find that a) Group B has significantly worse performance than Group A (CWS average: 13.1\% vs. 48.1\% ), b) Precision was hard to achieve for everyone (PR average: 43.8\%) but worse for Group B (0\%), c) Abstraction had moderately positive correlation to Decomposition and Precision for Group A (0.64), and d) we find that there was very little use of behavioral/dynamic modeling specification (BUML M- 0.38, CoV: 2.15) or formal notations for modeling/specification (NOTM M: 0.5, CoV: 1.79) (Table \ref{table:statscodes}).


\subsection{Group B has significantly worse performance than Group A}
As seen in Table \ref{table:statsscore}, Mean Composite Weighted Score for the entire set is 3.06 (out of 10), which is quite low. Group A (Tier 1 college students) has this at 4.81 and Group B (Tier 2/3 college interns at a company) at 1.31.  This suggests that Group B's performance has been significantly poorer compared to Group A (Table \ref{table:score}).  Precision similarly shows significant difference in performance between A and B (0.88 vs. 0).  In addition, Coefficient of Variance (CoV) for Group A lie between 0.53 and 0.95, while those for Group B are higher (1.18 to 1.38), which suggests more variability among the Group B students. 


\subsection{Precision (Completeness, Sufficiency) was hard to achieve for everyone}
Very few students (<22\%) were able to get the design to cater to all the scenarios; Group B didn't get any scenarios to pass.  Group A fared much better (Mean: 0.88). However, even for Group A, Coefficient of Variance (CoV) of 0.95 is high which suggests large variation among students.  Lack of precision can be either because of lack of design quality or it can be due to lack of expression quality.  A focus on a consistent structure of the documentation and being able to abstract well enough could help improve this score. 

\subsection{There wasn't enough expression diversity or density}
Using more expression types (diversity) like visual descriptions, flow charts, notations, Static UML, etc. and using more expressions in general (density) improves the quality of the description \cite{ajayi2009english,bezemer2008writing}. However, the documentation produced by the students used very few expression types (DV M:2.69,CoV:0.52), and very small number of expressions (DN M:4.69,CoV:0.58), resulting in a text-heavy and less clear documentation. We expected almost all the expression types to be used by all students.  

\subsection{There was a lack of formal notation usage for capturing dynamics of the design}
This was a surprising finding across the entire set of students.  Dynamics of the system (how the system moves from one state to another) is a key aspect of design and needs to be captured formally to improve precision, Li et al. \cite{li2006have} analyzed two large open-source software systems and showed that more than 80\% of defects are semantic (functionality related) in nature. However, less than 20\% of the expression usage included Behavioral UML or any set theoretic notations. Behavioral modeling aspect of UML was used quite less (BUML M: 0.38) and variation was large amongst participants (BUML CoF: 2.15). Same was the case for any notations/specifications (NOTM M: 0.5 CoF: 1.79). Not using these also caused lack of precision for many students. 

\subsection{Abstraction may be a key skill gap to fill}
A mean of 1 (out of 2) for Group A on Abstraction indicates that participants didn't do well. Correlation data (Table \ref{table:corrA}) shows abstraction has moderately positive correlation with Decomposition (0.64) as well as Precision (0.64) for Group A. This correlation was not expected and is worth investigating. If this correlation can point to causation, it is significant because it suggests that picking abstraction skills may help improving decomposition and modularization skills and may also make the design more precise.  

\begin{table}
\tiny
\caption{Pair-wise correlation for Group A scores}
\label{table:corrA}
\begin{tabular}{|l|r|r|r|} 
\hline
              & \multicolumn{1}{l|}{Decomposition}        & \multicolumn{1}{l|}{Abstraction}          & \multicolumn{1}{l|}{Precision}  \\ 
\hline
Decomposition & 1                                         & 0.64                                      & 0.44                            \\ 
\hline
Abstraction   & \textcolor[rgb]{0,0.502,0}{ 0.64} & 1                                         & 0.64                            \\ 
\hline
Precision     & \textcolor{blue}{0.44} & \textcolor[rgb]{0,0.502,0}{0.64} & 1                               \\
\hline
\end{tabular}

\end{table}

\begin{table}
\tiny
\centering
\caption{Design evaluation scores (N=16)}
\label{table:score}
\begin{tabular}{|c|c|c|c|c|c|c|c|} 

\hline
Name                                              & Grp                             & \textbf{DC} & \textbf{AB} & \textbf{PR} & \textbf{CWS} & \textbf{DV} & \textbf{DN}  \\ 
\hline
CP1                                               & A                                 & 1                                      & 1                                      & 1                                     & 5                                                  & 3                                                & 7                                            \\ 
\hline
CP2                                               & A                                 & 0                                      & 0                                      & 0                                     & 0                                                  & 1                                                & 1                                            \\ 
\hline
CP3                                               & A                                 & 2                                      & 1                                      & 0                                     & 3.5                                                & 2                                                & 3                                            \\ 
\hline
CP4                                               & A                                 & 2                                      & 2                                      & 2                                     & 10                                                 & 3                                                & 6                                            \\ 
\hline
CP5                                               & A                                 & 1                                      & 1                                      & 2                                     & 7.5                                                & 3                                                & 3                                            \\ 
\hline
CP6                                               & A                                 & 1                                      & 1                                      & 1                                     & 5                                                  & 5                                                & 9                                            \\ 
\hline
CP7                                               & A                                 & 0                                      & 1                                      & 0                                     & 1.5                                                & 4                                                & 7                                            \\ 
\hline
CP8                                               & A                                 & 2                                      & 1                                      & 1                                     & 6                                                  & 4                                                & 7                                            \\ 
\hline
ZP1                                               & B                                 & 0                                      & 0                                      & 0                                     & 0                                                  & 3                                                & 5                                            \\ 
\hline
{\cellcolor[rgb]{0.973,0.973,0.973}}ZP2           & B                                 & 1                                      & 0                                      & 0                                     & 1                                                  & 4                                                & 5                                            \\ 
\hline
ZP3                                               & B                                 & 2                                      & 1                                      & 0                                     & 3.5                                                & 1                                                & 2                                            \\ 
\hline
{\cellcolor[rgb]{0.973,0.973,0.973}}ZP4           & B                                 & 2                                      & 1                                      & 0                                     & 3.5                                                & 3                                                & 6                                            \\ 
\hline
ZP5                                               & B                                 & 0                                      & 0                                      & 0                                     & 0                                                  & 1                                                & 1                                            \\ 
\hline
{\cellcolor[rgb]{0.973,0.973,0.973}}ZP6           & B                                 & 0                                      & 0                                      & 0                                     & 0                                                  & 1                                                & 1                                            \\ 
\hline
ZP7                                               & B                                 & 1                                      & 1                                      & 0                                     & 2.5                                                & 3                                                & 5                                            \\ 
\hline
{\cellcolor[rgb]{0.973,0.973,0.973}}ZP8           & B                                 & 0                                      & 0                                      & 0                                     & 0                                                  & 4                                                & 8                                            \\ 
\hline
                                                  &                                   & \multicolumn{1}{l|}{}                  & \multicolumn{1}{l|}{}                  & \multicolumn{1}{l|}{}                 & \multicolumn{1}{l|}{}                              & \multicolumn{1}{l|}{}                            & \multicolumn{1}{l|}{}                        \\ 
\hline
Avg(all)                                     &                                   & 46.9\%                                 & 34.4\%                                 & 21.9\%                                & 30.6\%                                             & 2.8                                              & 4.8                                          \\ 
\hline
\textcolor[rgb]{0,0.502,0}{Avg(A)}     & \textcolor[rgb]{0,0.502,0}{A}     & \textcolor[rgb]{0,0.502,0}{56.3\%}     & \textcolor[rgb]{0,0.502,0}{50.0\%}     & \textcolor[rgb]{0,0.502,0}{43.8\%}    & \textcolor[rgb]{0,0.502,0}{48.1\%}                 & \textcolor[rgb]{0,0.502,0}{3.1}                  & \textcolor[rgb]{0,0.502,0}{5.4}              \\ 
\hline
\textcolor[rgb]{0.502,0,0.502}{Avg(B)} & \textcolor[rgb]{0.502,0,0.502}{B} & \textcolor[rgb]{0.502,0,0.502}{37.5\%} & \textcolor[rgb]{0.502,0,0.502}{18.8\%} & \textcolor[rgb]{0.502,0,0.502}{0.0\%} & \textcolor[rgb]{0.502,0,0.502}{13.1\%}             & \textcolor[rgb]{0.502,0,0.502}{2.5}              & \textcolor[rgb]{0.502,0,0.502}{4.1}          \\
\hline
\end{tabular}
\begin{description}
\item \textit{\textbf{DC}: Decomposition, \textbf{AB}: Abstraction, \textbf{PR}: Precision, \textbf{CWS}: Composite Weighted Score, \textbf{DV}: Expression Diversity, \textbf{DN}: Expression Density}
\end{description}
\arrayrulecolor{black}

\end{table}

\begin{table}
\tiny
\centering
\caption{Design Evaluation Summary Statistics (N=16)}
\label{table:statsscore}
\arrayrulecolor{black}
\begin{tabular}{|c|c|c|c|c|c|c|c|c|c|} 
\arrayrulecolor{black}\cline{1-1}\arrayrulecolor{black}\cline{2-10}
     & \multicolumn{3}{c!{\color{black}\vrule}}{All} & \multicolumn{3}{c!{\color{black}\vrule}}{Grp A} & \multicolumn{3}{c!{\color{black}\vrule}}{Grp B}  \\ 
\arrayrulecolor{black}\cline{1-1}\arrayrulecolor{black}\cline{2-10}
Name & M    & SD   & CoV                             & M    & SD   & CoV                               & M    & SD   & CoV                                \\ 
\hline
DC   & 0.94                                        & 0.85                                         & 0.91                                          & 1.13                                        & 0.83                                         & 0.74                                          & 0.75                                        & 0.89                                         & 1.18                                           \\ 
\hline
AB   & 0.69                                        & 0.60                                         & 0.88                                          & 1.00                                        & 0.53                                         & 0.53                                          & 0.38                                        & 0.52                                         & 1.38                                           \\ 
\hline
PR   & 0.44                                        & 0.73                                         & 1.66                                          & 0.88                                        & 0.83                                         & 0.95                                          & 0.00                                        & 0.00                                         & -    \\ 
\hline
CWS  & 3.06                                        & 3.04                                         & 0.99                                          & 4.81                                        & 3.20                                         & 0.66                                          & 1.31                                        & 1.60                                         & 1.22                                           \\ 
\hline
DV   & 2.69                                        & 1.40                                         & 0.52                                          & 3.00                                        & 1.51                                         & 0.50                                          & 2.38                                        & 1.30                                         & 0.55                                           \\ 
\hline
DN   & 4.69                                        & 2.73                                         & 0.58                                          & 5.25                                        & 2.96                                         & 0.56                                          & 4.13                                        & 2.53                                         & 0.61                                           \\
\hline
\end{tabular}
\arrayrulecolor{black}
\begin{description}
\item \textit{\textbf{DC}: Decomposition, \textbf{AB}: Abstraction, \textbf{PR}: Precision, \textbf{CWS}: Composite Weighted Score, \textbf{DV}: Expression Diversity, \textbf{DN}: Expression Density}
\item \textit{\textbf{M}: Mean, \textbf{SD}: Standard Deviation, \textbf{CoV}: Coefficient of Variance}
\end{description}
\end{table}

\begin{table}
\tiny
\centering
\caption{Expression Types Summary Statistics (N = 16)}
\label{table:statscodes}
\begin{tabular}{|l|p{0.2\textwidth}|r|r|r|} 
\hline
Code & Description                                                                            & \multicolumn{1}{l|}{M} & \multicolumn{1}{l|}{SD} & \multicolumn{1}{l|}{CoV}  \\ 
\hline
VDES & Visual Description (Diagrams)                                                          & 1.06                   & 1.24                    & 1.16                      \\ 
\hline
PROC & Procedures, Interfaces                                                                 & 0.25                   & 0.58                    & 2.31                      \\ 
\hline
ALDP & Pre-defined abstractions (algorithms, design patterns)                                 & 0.38                   & 0.50                    & 1.33                      \\ 
\hline
FLOW & Flow Chart or description                                                              & 0.81                   & 1.05                    & 1.29                      \\ 
\hline
SUML & Static modeling diagrams - UML or similar (Class, Object, Component)                   & 1.31                   & 1.45                    & 1.10                      \\ 
\hline
BUML & Behavioral/Dynamic modeling diagrams - UML or similar (Activity, Sequence, Statechart) & 0.38                   & 0.81                    & 2.15                      \\ 
\hline
NOTM & Modeling/Specification through notations (Z-notation, Sets/relations, English, etc.)   & 0.50                   & 0.89                    & 1.79                      \\
\hline
\end{tabular}
\begin{description}
\item \textit{\textbf{M}: Mean, \textbf{SD}: Standard Deviation, \textbf{CoV}: Coefficient of Variance}
\end{description}
\end{table}


\section{Conclusion and Limitations}
\label{sec:conclusions}

We evaluated 16 students' attempts to design a simple system and describe the design. Overall, the students demonstrated poor design and expression skills. Data suggests that most students found it hard to achieve precision in their design. Of those who did well in being precise, most of them also did well in abstracting their design.  Structure of these design documents were inconsistent, and this contributed to lack of precision as well, in addition to lack of usage of formal specification mechanisms like UML. This suggests lack of basic skills in technical documentation and articulation skills. These students came from two different college tiers and the difference between their performance in this experiment is significant. 

From teaching and learning perspective, there are a few areas where there should be focused intervention: a) Use abstraction well to achieve a good design, b) Use formal notations to achieve precision in the design, and c) Use defined structure and approach to produce good design documentation.

This also suggests why we have trouble producing employable engineers.  If these basic skills are missing, we can't expect these industry-ready students to do well in industry.  The contribution of this paper is in defining a way to evaluate these attributes and point towards areas to focus teaching and learning efforts in.  

\textit{Limitations}. These limitations of this study should be kept in mind when interpreting and using the results: a) This study was done on a small sample size (N=16), b) The difference between Group A and Group B performance can have many other factors confounding the results which were not controlled for (incentives may not be strong for Group B to perform better, for example, or the selection of students from these colleges.), c) The first author acted as the sole expert for analysis and this may have introduced a researcher's bias in the coding and scoring, and d) The rubric for expression quality evaluation is quite subjective. 

\textit{Future work}. We plan to run the experiment with more cohorts across different college categories and refine the encoding and rubric criteria. We also plan to identify curriculum interventions to enable students to acquire these skills effectively and efficiently. 

\textit{Applicability}. Even with this small sample size, a few key skill gaps in students are apparent and instructors can plan to focus on these: a) students struggle in abstracting well, b) students do not use enough diversity and density of expressions to articulate their design well, and c) students do not use enough formal notations of any kind.  Teachers can help students pick these skills through their course assignments.  In a subsequent paper, we plan to report on approaches to address these gaps through pedagogical interventions.

\bibliographystyle{ACM-Reference-Format}
\bibliography{compute22}

\end{document}